# Use Cases and Outlooks for Automatic Analytics

Joni Salminen & Bernard J. Jansen
{jsalminen,bjansen}@hbku.edu.qa
Qatar Computing Research Institute, Hamad Bin Khalifa University

**Abstract:** The landscape of analytics is changing rapidly. Much of online user analytics, however, is based on collection of various user analytics numbers. Understanding these numbers, and then relating them to higher numerical analysis for the evaluation of key performance indicators (KPIs) can be quite challenging, especially with large volumes of data. There is a plethora of tools and software packages that one can employ. However, these tools and packages require a quantitative competence and analytical sophistication that average end users often do not possess. Additionally, they often do little to reduce the complexity of numerical data in a manner that allows ease of use in decision making and communication. Dealing with numbers poses cognitive challenges for individuals who often do cannot recall many numbers at a time. Here, we explore the concept of automatic analytics by demonstrating use case examples and discussion on the current state and future of automated insights.

**Introduction:** For a long time already, automation has been invaded the field of marketing. Examples include scripts, rules and software taking care of pay-per-click optimization[1], machine learning optimizing display advertising (Google, 2017), tools generating automatic ad variations[2], and Web analytics software that tracks various information by default and alerting the users automatically. Regarding the latter, many advances toward automating analytics insights are currently being made in the industry and research fronts of data analytics. For example, there are several tools providing automated reporting functions (e.g., Google Analytics, TenScores, Quill Engage, Qwaya, etc.).

While some of these tools require pre-configuration such as creating report templates, it is becoming more common that the tool itself chooses the relevant insights it wants to portray, and then delivers these via email to decision makers. An example of such an approach is provided in Figure 1, which shows Quill Engage, a tool that automatically creates fluent text reports from Google Analytics data.

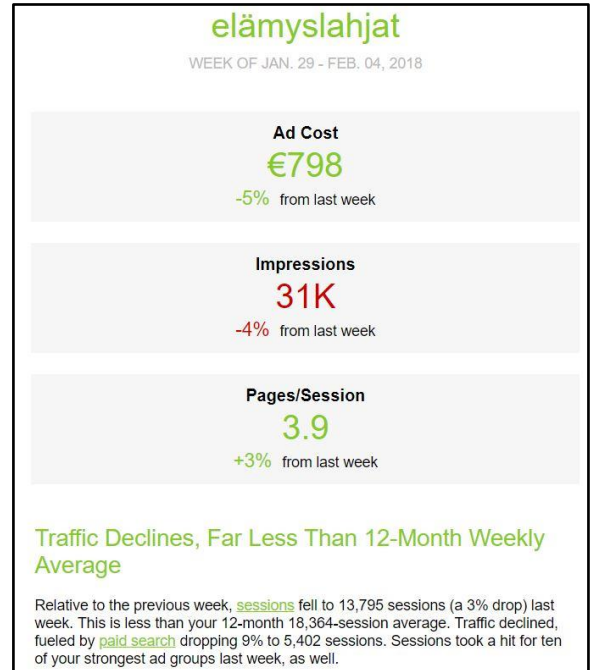

**Figure 1: Quill Engage. The Tool Automatically Generates Fluent Reports from Google Analytics Data, And Provides Numerical Comparisons Based on Outliers and Trends.**

As can be seen from Figure 1, the automatic analytics tool quickly displays key information and then aims to provide *context* to it that provides potential reasoning to the key performance indicators. The benefits of automatic analytics are obvious. First of all, they spare decision makers' time, as they are not forced to log into systems, but receive the insights conveniently to their email inboxes and can rapidly take action. Cognitive limitations (Tversky & Kahneman, 1974) are imposing serious constraints for decision makers dealing with ever-increasing amounts of "big data," highlighting the need for smart tools that pre-process and mine the data at the user's convenience.

The issue that automatic analytics is solving is complexity: as a marketing manager, one has many platforms to manage and many campaigns to run within each platform. In sum, several data sources and metrics introduce such a degree of complexity that it becomes impossible for human beings, constrained by limitations of cognitive capacity, to make sense of the data.

---

[1] https://www.linkedin.com/feed/update/urn:li:activity:6367455477760163840

[2] https://www.qwaya.com/tour

**Previous work:** To accomplish the goal of easy-to-use analytics, one popular area of advanced analytics is natural language systems, where users find the information by asking the system in free format. For example, previously, Google Analytics had a feature called Intelligence Events, which detected anomalies, but this was removed recently. Currently, Google provides automatic insights via a mobile app, in which the user can ask the system in natural language to provide information. An example of this is provided in Figure 2.

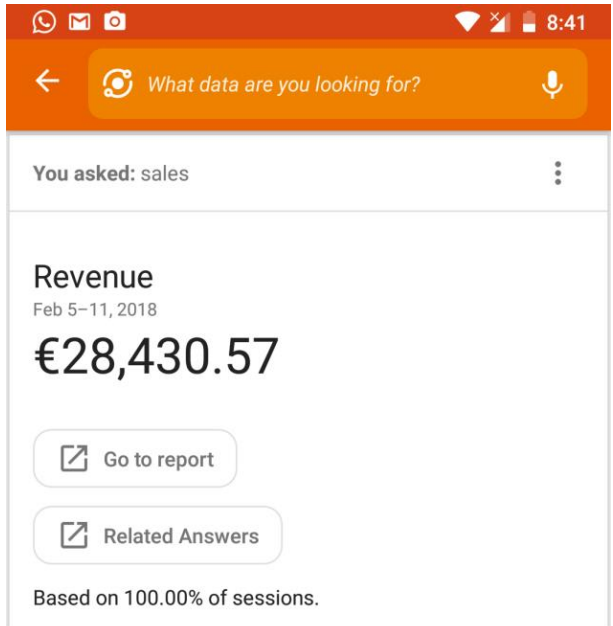

**Figure 2: Screenshot from Google Analytics Android App, Showing the Functionality of Asking Questions from the Analytics System.**

However, even asking the system requires effort and prior knowledge. For example, what if the question is not relevant or misses an important trend in the data? For such cases, the system must anticipate, and in fact analyze the data beforehand, instead of only provide reports for human analysts. This is a central feature in automatic analytics systems.

We conceptualize this issue as having two primary use cases for data analytics: (1) deep analyses and (2) day-to-day decision making. While one periodically needs to perform a deep analysis on strategic matters, such as updating online marketing strategy, creating a new website structure, etc., the daily decisions cannot afford a thorough use of tens of reports and hundreds of potential metrics. That is why many reports and metrics are not used by decision makers in the industry at all. The solution to this condition has to be automation: the systems have to direct human users' attention toward noteworthy things. This means detecting anomalies on marketing performance, predicting their impact and presenting them in actionable format to decision makers, preferably by pinging them via email or other channels, such as SMS. The systems could even directly create tasks and push them to project management applications like Trello. A requisite to automatic analytics should therefore be the well-known SMART formula, meaning that Specific, Measurable, Appropriate, Realistic and Timely goals (Shahin & Mahbod, 2007). Through this principle, decision makers can rapidly turn insights into action.

**Use cases:** In the following, we provide some examples of current state-of-the-art tools of automatic analytics. We then generalize some principles and guidelines based on an overview of these tools. First, in Figure 3, we present a screenshot from email sent by TenScores, a tool that automatically scans Quality Scores for Google AdWords campaigns.

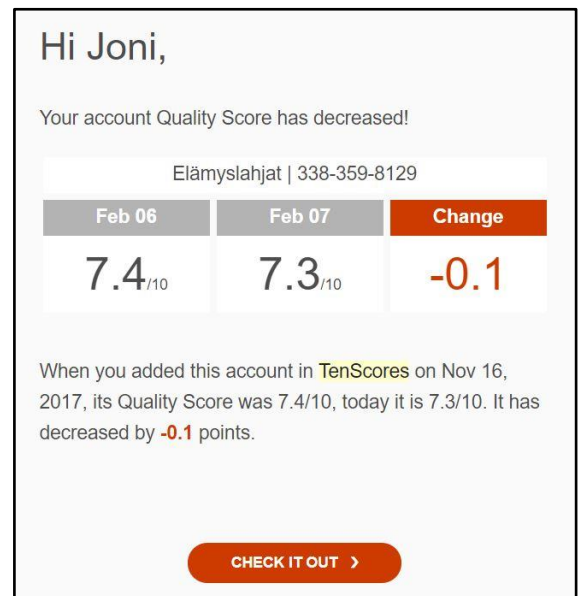

**Figure 3: TenScores, the Automatic Quality Score Monitoring Tool.**

In search-engine advertising, Quality Scores are important because they influence the click prices paid by the advertisers (Jansen & Schuster, 2011; Salminen, 2009). In this case, the tool informs when there is a change in the average Quality Score of the account. From a user experience perspective, the threshold to alerting the user is set to very low change, resulting in many emails sent to the users. This highlights the risk of automation becoming "spammy," leading into losing user interest. The correct threshold should be set according to open rates by experimenting with different increments.

In Figure 4, we can see a popular Finnish online marketplace, Tori.fi. Tori sends automatic emails to its corporate clients, showing how their listings have performed compared to previous period, and enabling the corporate clients to take direct action from within the email. From example, one can click the blue button and the particularly listing which is not performing well, is boosted. In addition, there is a separate section (not visible from the screenshot) showing the best performing listings.

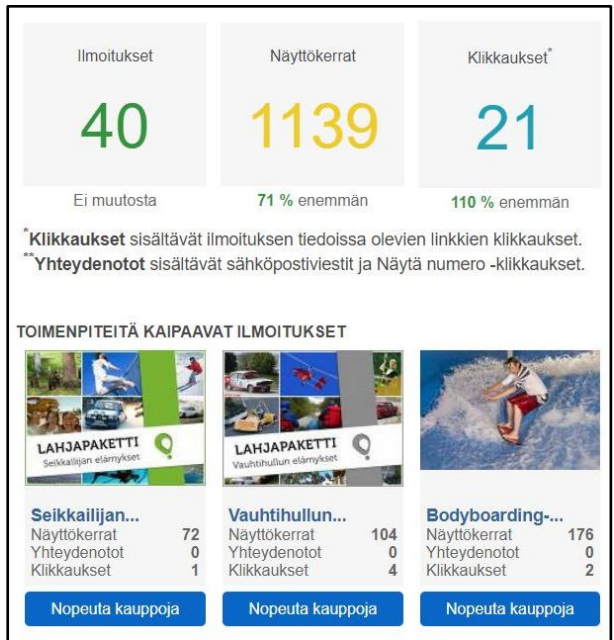

**Figure 4: Tori's Marketplace Insights Automatically Delivered to Inbox.**

**Risks:** However, there are not only opportunities associated with automatic analytics, but also risks. For example, in search-engine advertising, brands are bidding against one another (Jansen & Schuster, 2011). Thus, an obvious step to further optimize their revenue by providing transparent auction information is Google sending automatic emails when the relative position (i.e., competitiveness) of a brand decreases, prompting advertisers to take action. This potential scenario also raises questions about morality and ethics of automated analytics, especially in click auctions where the platform owners have an incentive to recommend actions that inflate click prices (Salminen, 2009). For example, in another online advertising platform, Bing Ads, the "Opportunities" feature gives suggestions marketers can implement in a click of a button. However, many of these suggestions relate to increasing the bid prices (see Figure 5).

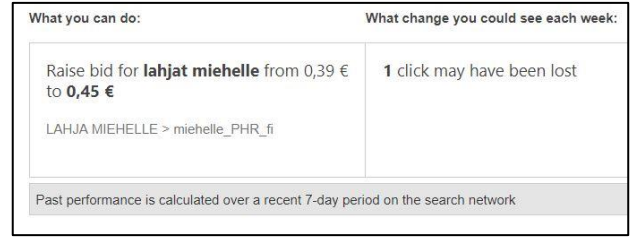

**Figure 5: An example of Bing Ads Recommending to Increase Keyword Bid.**

If the default recommendation is always to raise bids, the feature does not add value to end user but might in fact destroy it. From an end user point of view, therefore, managers are encouraged to take recommendations with a grain of salt in such cases. From a research point of view, it is an interesting question to find out how much the automatic recommendations drive user actions.

**Discussion:** To conclude, we present a few visions about the future of automatic analytics, especially in relation to online marketing as the tools in that are among the most developed ones. First off, the current situation in online marketing is that optimization consist of hundreds of micro-tasks that are interconnected and require analytics skills and creativity to be solved in a close-to-optimal way. We see the role of automated analytics as pre-filtering this space of potential tasks into a manageable number to human decision makers and assigning the tasks a priority number. In the end, the human decides which ones to act upon, but this automatic filtering and sorting is highly beneficial in maneuvering the fragmented channel and campaign landscape currently taking place in online marketing.

Moreover, each vertical has its own KPIs, metrics and questions, resulting in requirement of many tools. For example, search-engine optimizers require drastically different information than display advertisers, and therefore it makes no sense to create "one size fits all" solution where the information needs of the users are so different. Instead, an organization should derive the tools from its business objectives and based on the specific information needed to achieve them, as presented by Järvinen and Karjaluoto (2015). An example of such fine-grained automatic analytics is TenScores that only specializes on monitoring one metric in one channel (Quality Score in Google AdWords). Their approach makes sense because Quality Score is such an important metric for keyword advertisers and its optimization involves a complexity, enabling TenScores to provide in-depth recommendations valuable to end users.

However, even though the tools may be channel-specific, their operating principles can share a great deal of similarity. For example, stream filtering and anomaly detection algorithms are generalizable to many types of data, and thus have great applicability in practical system implementations. In a similar vein, setting the threshold to emails is a key issue that can be experimented with when designing these systems. When rules need to be defined manually, the cognitive limitations further steer away from reaching optimal solutions because the capacity of individuals to capture the optimal parameter values from a vast range of possible values almost certainly results in sub-optimality. Algorithmic approaches, such as greedy multi-armed bandit optimization (Leet el al., 2013), are much more efficient in dealing with large arrays of choices requiring constant exploration. Therefore, even if there is automation, it is too early to speak of real artificial intelligence. The current systems always have manually set parameters and thresholds and miss important things that are clear for individuals. For example, the previously shown Quill Engage cannot provide an explanation why the sales dropped when going from December to January – but this is apparent to any individual working in the gift business: Christmas sales were the reason.

In conclusion, developers of various analytics systems should no longer expect that their users log-in to the system to browse reports, but the critical information needs to be automatically mined and sent to them in a format that enables precise actions (cf. SMART principle). There is already a considerable shift in the industry to this direction which will only be emphasized as customers realize the benefits of automatic analytics. Thus, we believe the future of analytics is more about detecting anomalies and opportunities and giving decision makers easy choices to act upon. Of course, you get also new concerns in this environment, meaning that the recommendations can be purposefully or un-purposefully biased. For example, there's a risk of a moral hazard in Google's recommendations – are they advising you to increase bids because it maximizes your profit or because it increases their revenue? And, if they suggest two advertisers competing over the same keywords to raise their bids, even either one taking the suggestion can lead into escalating bids.

**Conclusion:** Analytics software providers are planning to move toward the direction of providing automated insights, and researchers should follow suite. Open questions are myriad, especially relating to users' interaction with automatic analytics insights: how responsive are they? What information do they require? What actions do they take based on the information? We expect interesting studies in this field in the near future.